\documentclass[journal]{IEEEtran}
\usepackage{amsmath}
\usepackage{algorithm}
\usepackage{algorithmicx}
\usepackage{algpseudocode}
\usepackage{amssymb}
\usepackage{array}
\usepackage{cite}

\ifCLASSINFOpdf
   \usepackage[pdftex]{graphicx}
   \graphicspath{{../pdf/}{../jpeg/}}
   \DeclareGraphicsExtensions{.pdf,.jpeg,.png}
\else
   \usepackage[dvips]{graphicx}
   \graphicspath{{../eps/}}
   \DeclareGraphicsExtensions{.eps}
\fi

\begin{document}

\title{Polarized Low-Density Parity-Check Codes \\on the BSC}

\author{Binbin~Gao$^1$,
        Huarui~Yin$^1$,~\IEEEmembership{Member,~IEEE}\\
        Zhengdao~Wang$^2$,~\IEEEmembership{Fellow,~IEEE}
        \quad \\
        $^{1}$ CAS Key Laboratory of Wireless-Optical Communications,\\
    	University of Science and Technology of China, Hefei, P.R.China\\
    	$^{2}$ Department of Electrical and Computer Engineering,\\
    	Iowa State University, Ames IA, USA 50010\\
Email: gbb9505@mail.ustc.edu.cn, yhr@ustc.edu.cn, zhengdao@iastate.edu}

\maketitle

\begin{abstract}
The connections between variable nodes and check nodes have a great influence
on the performance of low-density parity-check (LDPC) codes. Inspired by the
unique structure of polar code's generator matrix, we proposed a new method of
constructing LDPC codes that achieves a polarization effect. The new code,
named as polarized LDPC codes, is shown to achieve lower or no error floor in the
binary symmetric channel (BSC).
\end{abstract}


\IEEEpeerreviewmaketitle

\section{Introduction}

\IEEEPARstart{L}{ow}-density parity-check (LDPC) codes, first discovered by Gallager
\cite{gallager1962low}, had not been used in practice for several decades
due to lack of efficient decoding algorithms. It was rediscovered by Luby and
MacKay et al.
\cite{luby1998analysis,mackay1999good}. Thanks to LDPC’s
capacity-approaching performance and low iterative decoding complexity, it has
been applied in many wireless communication systems, e.g., in the Digital
Video Broadcasting Satellite - Second Generation (DVB-S.2) Standard. With the
efforts from both academia and industry, LDPC codes have also been adopted as an
eMBB traffic-channel coding scheme of the fifth generation wireless
communications.

To trace the performance of iterative decoding, Richardson and Urbanke
\cite{richardson2001capacity} proposed an algorithm called Density Evolution
(DE) to calculate the probability density function of variable nodes and check
nodes in each iteration. The DE algorithm shows that LDPC codes with a certain
degree distribution would have an arbitrarily small bit-error rate (BER) when
the code length tends to infinity and the level of channel noise is below a
threshold. Otherwise, the BER would be larger than a positive constant.
Density evolution is useful for finding a theoretically good degree
distribution, which is fundamental for the construction of practical LDPC
codes. With the DE algorithm, Chung et al.\ found a rate-1/2 code with a good
degree distribution that achieved 0.0045 dB within the Shannon limit for
additive white Gaussian noise (AWGN) channel \cite{chung2001design}.

One of the most important assumption of density evolution algorithm is the
local-tree assumption, namely, the subgraph generated after the $l$-th iteration
remains a tree. However, realistic codes usually have cycles in the Tanner
graph representation which would render the assumption invalid after
sufficient number of iterations. Intuitively, the cycles, especially those
with small girths, obstruct the flow of extrinsic information among the nodes.
There are several successful algorithms to construct large-girth LDPC codes.
One such algorithm is the Tanner-graph based on progressive-edge-growth (PEG)
algorithm proposed by Hu et al.~\cite{hu2005regular}, which aims to maximize
the minimum girth. However, girth is not the only factor affecting the
performance of LDPC codes. In \cite{tian2004selective}, Tian et al.\ pointed
out that the connectivity among nodes is also important. The extrinsic message
degree (EMD) measures variable node connectivity in the bipartite graph of
LDPC codes. The approximate cycle EMD (ACE) is defined as the upper bound on
the EMD of all variable nodes in a given cycle. Combining the PEG and the ACE
algorithm, Xiao and Banihashemi proposed an improved PEG algorithm
\cite{xiao2004improved}.

Although LDPC codes have capacity approaching performance, it has been proved
that LDPC codes cannot reach the Shannon limit without infinite variable degree
even with infinite code length.

Polar codes, proposed by Arikan \cite{arikan2008channel}, have been
mathematically proved to be able to achieve the Shannon limit on the binary
symmetric channel (BSC) with infinite code length. Compared to LDPC codes,
polar codes have better error-correcting performance in the short code length
or low rate situations.

However, compared with LDPC codes, polar codes have higher decoding
complexity, due to the fact that the parity check matrix is not sparse.

It is an interesting question whether we could combine LDPC codes with polar
codes. 
%Inspired by the concatenated coding \cite{forney1965}, it has been
%proposed to concatenate a polar code as inner code with an LDPC code as outer code
%in \cite{eslami2013finite}. 
Inspired by the concatenated coding \cite{forney1965}, it has been
proposed to concatenate a polar code as outer code with an LDPC code as inner code in \cite{eslami2013finite}.
For practical encoding over finite length, the
ideally polarized channels of polar code are only semi-polarized. For these
cases, it may be a good way to use LDPC codes to further protect the bits
transferred on the channels \cite{yu2018improved}. However, the concatenation
of polar codes and LDPC codes does not address the problem of high decoding
complexity of polar codes. A practical low-complexity soft decoding algorithm
for polar codes remains yet to be found. Instead of improving polar codes with
LDPC codes, it may be a good idea to improve LDPC codes with polar codes.

In this paper, we propose a new method of constructing LDPC codes with
inspiration from polar codes. Through judicious placement of the edges
connecting the variable and parity-check nodes, we achieve polarization of the
variable and parity-check nodes. With slight increase of decoding complexity,
the new code enjoys lower BER and faster convergence on the BSC. Also we have
not observed error floor in our simulated cases, which is a known problem for
conventional LDPC codes.

The organization of the paper is as follows. In the next section, we review the basic concept like Tanner graph and degree polynomials in LDPC codes. Inspired by polar codes, we propose a new method constructing LDPC codes called polarized LDPC codes. And the improvements on density evolution and decoding algorithm are discussed. In Section III, using improved PEG algorithm, we construct realistic polarized LDPC codes and present simulation results comparing the performance of polarized and standard LDPC codes in the binary symmetric channel. Finally some conclusions are drawn in Section IV.

\section{Polarized LDPC}

Tanner graph, proposed by Tanner \cite{tanner1981recursive} in 1981, with
original purpose of constructing long error-correcting codes from sub-codes,
is one of the most important tools describing LDPC codes. A Tanner graph is a
bipartite graph. One type of nodes is called variable nodes representing
the bits in the codeword. The other type is called check nodes, representing
the parity-check equations. The edges between variable nodes and check nodes
represent the coded bits that a check equation involves.

The number of edges linked to a node is defined as the degree of the node.
Tanner graph is in accordance with parity-check matrix $H$. A variable node
corresponds to a column of the matrix, and a check node corresponds to a row
of the matrix. If there is an edge between the $j$-th check node and $i$-th
variable node, the $(j,i)$-th element of $H$ is set to 1. A random irregular
LDPC code can be defined by two degree polynomials:
\begin{equation}
\lambda(x)=\sum_{i=2}^{d_v}\lambda_i x^{i-1}
\end{equation}
and
\begin{equation}
\rho(x)=\sum_{j=2}^{d_c}\rho_j x^{j-1}
\end{equation}
where $\lambda(x)$ is variable degree polynomial and $\rho(x)$ is check
degree polynomial, $\lambda_i$ and $\rho_j$ represent the fraction of edges
connected to degree-$i$ variable nodes and those connected to degree-$j$ check
nodes, respectively. Viewed from another perspective, $\lambda_i$ can be
interpreted as the probability of any check node having a common edge with a
degree-$i$ variable node, and $\rho_j$ is the probability of any variable node
having a common edge with a degree-$j$ check node.

\def\vij{v_{i\to j}}
\def\uji{u_{j\to i}}

Given a received symbol $y$ and a transmitted bit $x$, we define log-likelihood
ratio (LLR) in the form of
\begin{equation}
  \log\frac{P(y|x=0)}{P(y|x=1)}.
\end{equation}
Let $\vij$ be the LLR message from variable node $i$ to check node $j$. And
$\uji$ be the LLR message from the check node $j$ to variable node $i$. Let
$v_{i,0}$ be the LLR message from the channel. Let $U_i$ denote the set of
check nodes that are connected to variable node $i$. Similarly, define $V_j$
as the set of variable nodes connected to check node $j$. According to the
sum-product decoding algorithm, e.g.,
\cite{richardson2001capacity}, $\vij$ is updated by:
\begin{equation}\label{eq.rule1}
\vij = v_{i,0} + \sum_{j'\in U_i, j'\ne j}u_{j'\to i}
\end{equation}
and $\uji$ is updated by the ``tanh'' rule:
\begin{equation}\label{eq.rule2}
\tanh \left(\frac{\uji}{2}\right) =
 \prod_{i'\in V_j, i'\ne i} \tanh \left(\frac{v_{i'\to j}}{2}\right).
\end{equation}

Define the $R$-calculation \cite{chung2001design} as
\begin{equation}
  R(a,b)= 2\tanh^{-1}\left (\tanh\left(\frac a2\right) \tanh \left(\frac
  b2\right) \right).
\end{equation}
We can rewrite a ``tanh'' rule such as
\begin{equation}
\tanh\left(\frac u2\right)= \prod_{i=1}^{d-1} \tanh\left(\frac {v_i}2\right)
\end{equation}
as
\begin{equation}
u=R(v_1,R(v_2,...,R(v_{d-2},v_{d-1}))).
\end{equation}

If we view the LLR messages as random variables due to the stochastic nature
of the channel, based on the sum-product decoding rules \eqref{eq.rule1} and
\eqref{eq.rule2}, the corresponding rules of the transformation on the
probability density functions (PDF) of LLR can be derived. The update rule for
the PDF of LLR on the variable node side is basically convolution, and the
update rule for PDF on the check node side can be expressed in a form similar
to convolution; for details see e.g., \cite{chung2001design}.

\subsection{LDPC polarization: observation}

Assuming that all-0 word is sent and the crossover probability of BSC is
$\epsilon<1/2$, the initial density function of $\{v_{i,0}$ for any $i$ is
\begin{equation}
p_0(x) \!=\!\epsilon \delta\left(\!x\!+\log(\frac{1\!-\!\epsilon}{\epsilon}\!)\right)\!+\!
  (1\!-\!\epsilon) \delta\left(\!x\!-\!\log(\!\frac{1\!-\!\epsilon}{\epsilon}\!)\right).
\end{equation}

We have the following important observation on the update rules
\eqref{eq.rule1} and \eqref{eq.rule2}. Under the assumption of all-0 codeword,
adding an extra edge to a neighborhood set $U_i$ will tend to increase the LLR
$\vij$. On the other hand, because $\tanh(x)\in (-1,1), \forall x$, removing
an edge from the neighborhood set $V_j$ will tend to increase the LLR $\uji$.
Loosely speaking, variable nodes with higher degrees are more reliable. On the other hand, check nodes with lower degrees are more reliable. Thus, if we connect higher-degree variable nodes with lower-degree check nodes, and
connect lower-degree variable nodes and higher degree check nodes, we will
create a polarization effect: higher-degree variables nodes are more reliable and lower-degree ones are less reliable.

In standard LDPC codes, under the restriction of degree polynomials, the
edges between variable nodes and check nodes are established randomly. In
contrast to this random connectivity, polar codes have connections that are
structured and polarized. Taking a rank-4 polar matrix $G_4$ as an example:
\begin{equation}
  G_4 = \left[ \begin{matrix}
      1& 0& 0& 0\\1& 0& 1& 0\\1& 1& 0& 0\\1& 1& 1& 1\\
      \end{matrix}.
    \right]
\end{equation}
Unlike the H matrix of LDPC code, the columns of $G_4$ can be thought of as
check nodes (the polarized channel), while the rows as variable nodes (the
original channel). The polarized channel with higher capacity has lower
``check'' degree, while the original channel having higher ``variable''
degree has more probability to correct errors. Using serial interference
cancellation, the decoding procedure would decode first the most reliable
bits, cancel their interference, and then detect the weaker bits with the
prior information.

A polarized LDPC code may be preferable for the decoding performance and
complexity. Intuitively, polarized bits that are reliable can be stabilized
and decoded quickly, which is helpful to cancel their interference on the
check nodes connected to them. This in turn is helpful for decoding other
variable bits with lower reliability. Such benefits of polarization are well
established in polar codes and support our following polarized LDPC design.

\subsection{LDPC polarization: code construction}

\begin{figure}[!t]
	\centering
	\includegraphics[width=0.5\textwidth]{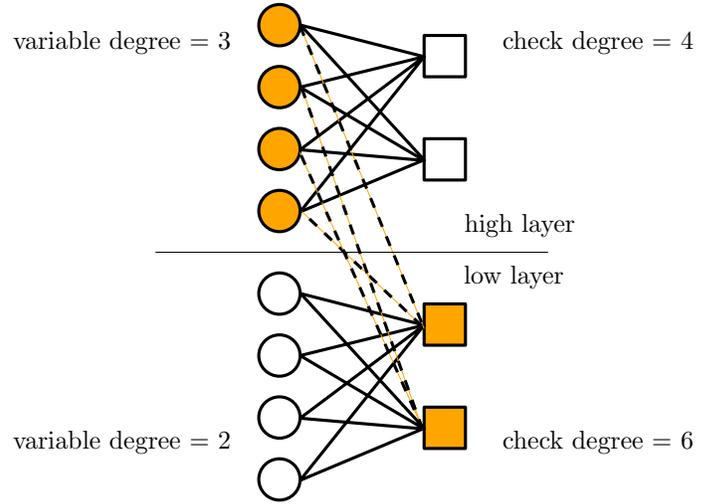}
	\caption{Tanner Graph of polarized LDPC code}
	\label{fig_TG}
\end{figure}

A polarized LDPC code can be constructed by connecting low-degree variable nodes to high-degree check nodes, and low-degree check nodes to high-degree variable nodes. Fig.~\ref{fig_TG} gives an example of polarized LDPC codes. We divide the nodes into two layers. The high layer contains variable nodes with higher degree (degree 3) and check nodes with lower degree (degree 4). The low layer contains degree-2 variable nodes and degree-6 check nodes. We have used dashed lines to indicate the connections between the high-degree variable nodes and high-degree check nodes (the inter-layer connections).

In general, we divide both variable nodes and check nodes into layers, with
the degree of variable nodes ordered in descending order from top layer to
bottom layer, and the degree of check nodes ordered in ascending order. We
connect variable nodes in a layer to check nodes in the same layer and all
layers below it.

In standard LDPC codes, variable nodes and check nodes are treated as two
different sets because the connectivity between variable nodes and check nodes
is independent of their degrees. So the node in the same set would get the same
information. For example, the output LLR of variable nodes
$v=\sum_{i=2}^{d_v}\lambda_i v_i$ would be the incoming LLR for any degree
check nodes. But in polarized LDPC codes, variable nodes with different degrees must
be treated as different sets because they would have different probabilities
to have connections with degree-$j$ check node for any given $j$. And so do
the check nodes. Then nodes with different degrees would have different degree
polynomials. Let $\rho_i(x)$ denote the check degree polynomial of a variable
node of degree $i$, and $\lambda_j(x)$ the variable degree polynomial of a
check node of degree $j$. Different from standard LDPC codes, our polarized
LDPC codes design would create layers having different rate according to the degree
of variable nodes, in a way similar to that of the polar code. Different layers
of variable nodes would exchange information by their common check nodes,
through the inter-layer connections, as indicated by dashed lines in the
example in Fig.~\ref{fig_TG}.
%\figref{fig_TG}.

After the higher degree variable nodes are decoded, the LLR sent to the common
check nodes would be set to infinity (assuming all 0-codeword is sent). So the
next layer variable nodes could cancel the interference in the decoding
procedure to get more capability to correct errors. Because traditional
methods of analyzing LDPC codes are based on the random link between nodes, we
proposed an improved density evolution analysis method to analyze the
polarized LDPC codes; see Algorithm~\ref{algo}.

\begin{algorithm}[ht]
	\caption{Polarized Density Evolution}\label{algo}
	\begin{algorithmic}[1]
		\State Calculate degree-i variable node's check degree is: $\rho_i(x)$
		\State Calculate degree-j check node's variable degree is: $\lambda_j(x)$
		\For {iter=1:iternum}

			\For {$i=2:d_v$}
				\State The input PMF of LLR from linked check nodes: $p^{(l)}_{u_i}=\sum_{j=2}^{d_c}\rho_{ij} p^{(l)}_{u_j}$
				\State The output PMF of LLR of degree-i variable node: $p^{(l+1)}_{v_i} = p_{u_0}*\otimes^{i-1}(p^{(l)}_{u_i})$
			\EndFor

			\If{$p^{(l+1)}_{v_i}$ tends to the ``point mass at infinity''}
			\State Change the degree-i variable node's check degree: $\rho_i(x)$ by cutting the edges to the higher degree check nodes.
			\EndIf

			\For {$j=2:d_c$}
				\State The input PMF of LLR from linked variable nodes: $p^{(l+1)}_{v_j}=\sum_{i=2}^{d_v}\lambda_{ji} p^{(l+1)}_{v_i}$
				\State The output PMF of LLR of degree-j check node: $p^{(l+1)}_{u_j} = R^{j-1} p^{(l+1)}_{v_j}$
			\EndFor

			\State The PMF of LLR of check node is: $p_u^{(l+1)} = \sum_{j=2}^{d_c}\rho_jp^{(l+1)}_{u_j}$

			\If{$p_u^{(l+1)}$ tends to the ``point mass at infinity''} \\
      Stop the iteration, all the nodes have been corrected.
			\EndIf
		\EndFor
	\end{algorithmic}
\end{algorithm}
In the standard DE algorithm, all output LLR of variable/check nodes are
assumed to be independent and identically distributed (i.i.d) variable, which
is not true in the polarized LDPC. The first step of our algorithm is
calculating the check degree polynomial $\rho_i(x)$ of degree-$i$ variable
node and regarding the coefficient as $\rho_{ij}$, which expresses the
probability of the degree-$i$ variable node having connection with the
degree-$j$ check node. We calculate degree-j check node's variable degree
polynomial $\lambda_j(x)$ and regard the coefficient as $\lambda_{ji}$ in the
same way. In the iterations, each node with a different degree would have a
unique LLR mixture as input and a different output, which are calculated and
stored respectively.

\subsection{LDPC polarization: decoding}

The decoding procedure of polarized LDPC codes proceeds layer by layer. The higher
layer bits would have errors corrected in a few iterations while the lower
layer bits need more iterations. When the probability of error of high layer
tends to zero, the information passed from the lower degree variable nodes
which might have uncorrected bit would interfere with the already corrected
bits. So after the density of the degree-$i$ variable nodes tending to the
``point mass at infinity'', which means the probability of error tends to
zero, we would change $\rho_i(x)$ making it only pass information to the
uncorrected nodes instead of receiving information from them. However, this
modification of the standard decoding procedure is not critical.

\begin{figure}[!t]
	\centering
	\includegraphics[width=0.5\textwidth]{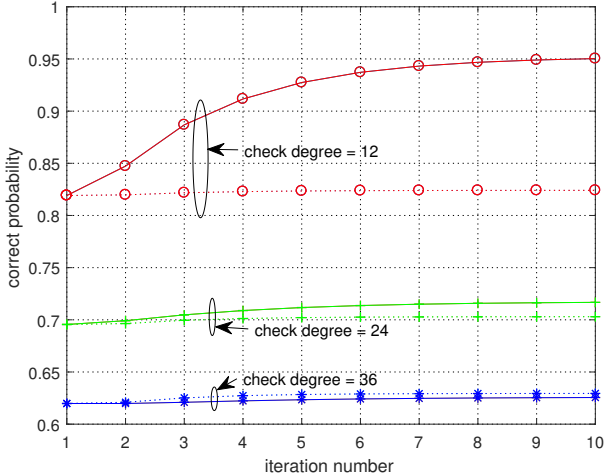}
	\caption{DE based calculated correct decoding probabilities of
  check nodes of different degrees:
  Standard LDPC (dotted lines) and Polarized LDPC (solid lines)}
	\label{fig_DE}
\end{figure}

To see the effect of polarization, we use an LDPC code with variable degree
polynomial $\lambda(x)=0.0417x+0.8333x^3+0.1250x^5$ and check degree
polynomial $\rho(x)=0.0417x^{11}+0.8333x^{23}+0.1250x^{35}$, and construct the
bipartite graph using three layers according to our polarized code
construction rules. Assuming that the all-0 word is sent, Fig.~\ref{fig_DE}
correct probability of different degree check nodes with polarized and
standard density evolution in 10 iterations for BSC.
In this simulation, the crossover probability is higher than the correct-decoding threshold.
The correct probability of standard LDPC code remains constant with iterations which is consistent with the threshold phenomenon of LDPC.

The simulation result indicates that the check nodes in standard LDPC codes behave
similarly due to the random connection nature. For polarized LDPC codes, the
degree-12 check node curve has a larger growth rate than the other two curves
thanks to the structured nature of polarized connections. As a result,
polarized LDPC codes can be decoded layer by layer. The lower-degree check
nodes get more reliable LLR from the higher-degree variable nodes, so the high
layer would get the greatest gain which is the aim when designing
polarized LDPC codes. So using the same degree polynomials, polarized LDPC codes
can decode part of information bits instead of throwing away all the code word
like in standard LDPC codes. This property of partial decoding can be useful for
certain applications.

\section{Simulation Results}

In this section, we construct realistic codes to check the performance on the
BSC. We choose the Improved Progressive Edge Growth (IPEG) algorithm
\cite{xiao2004improved} to construct polarized LDPC codes. Different from the
original IPEG algorithm, to generate polarized LDPC codes, the variable node
degree sequence should be in non-increasing order while the check node
degree sequence should be in non-decreasing order. So the edges between
high-degree variable nodes and the low-degree check nodes would be generated
first. The IPEG algorithm would make connections from variable nodes in the higher
layer to check nodes in neighbor layers as much as possible.

Using the IPEG algorithm, we construct four irregular polarized LDPC codes
with the same code length, rate, and the same numerical value of variable and
check degrees. All four codes' length are equal to 16384, the code rate equals
5/6. The only difference among the four LDPC codes is the coefficients of
their variable and check degree polynomials, which decide the correlation
between layers. Table~\ref{table_example} gives the coefficients of the degree
polynomials. We simulate on the BSC with iterative belief propagation
decoding. The maximal number of iterations is 50. And there are $10^5$ frames
for each crossover probability simulated.

\begin{table}[!h]
\renewcommand{\arraystretch}{1.3}
\caption{The Degree Distribution Table}
\label{table_example} \centering
\begin{tabular}{|c||c|c|c|}
\hline Variable/Check degree value&2/12&4/24&6/36\\ \hline code
A&0.0417&0.8333&0.1250\\ \hline code B&0.05&0.8&0.15\\ \hline code
C&0.0625&0.75&0.1875\\ \hline code D&0.0833&0.6667&0.25\\ \hline
\end{tabular}
\end{table}

\begin{figure}[!t]
	\centering
	\includegraphics[width=0.5\textwidth]{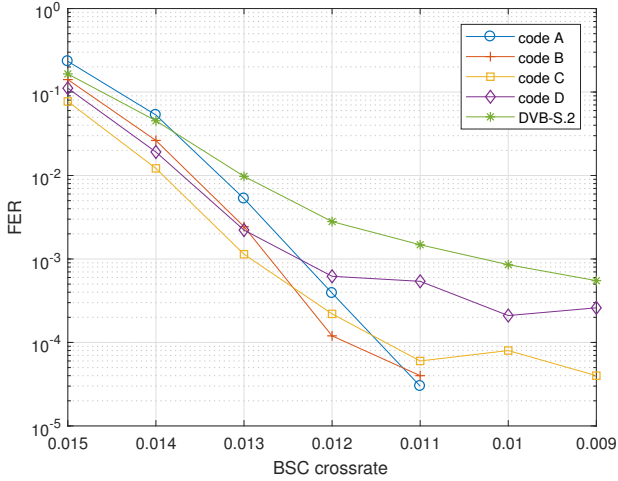}
	\caption{The FER of four different polarized LDPC codes and DVB-S.2 LDPC codes with code rate=5/6}
	\label{fig_FER}
\end{figure}

Fig.~\ref{fig_FER} gives the FER performance of four polarized LDPC codes and
DVB-S.2 short LDPC code. In the high crossover probability region, the
polarized LDPC codes and the DVB-S.2 code have close performance. When it
comes to low crossover probability region, the DVB-S.2 LDPC code has a clear
error-floor phenomenon. However, the FER of code C and code D have a much lower
error-floor, while code A and code B continue decreasing to zero. We
also note that the coefficients of degree polynomial have great influence on
the code performance. It seems a smaller percentage of low-degree variable
nodes would lead to a lower error-floor while a larger percentage of low-degree
nodes may have some advantage in the high crossover probability region. The exact effect of coefficients of degree polynomials on the code performance
deserves further investigation.

\begin{figure}[!t]
	\centering
	\includegraphics[width=0.5\textwidth]{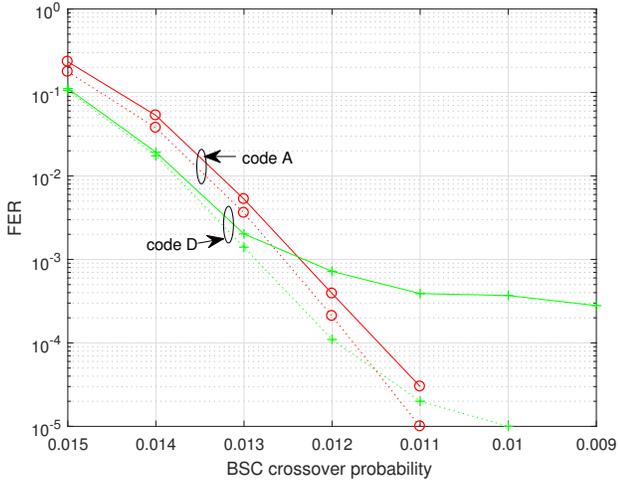}
	\caption{The FER of variable nodes of two different degree polarized LDPC codes:
  degree 4 (solid) and degree 6 (dotted) }
	\label{fig_degree_FER}
\end{figure}

Fig.~\ref{fig_degree_FER} gives the FER performance of different layers of
polarized LDPC codes. As we can see, the degree-4 variable nodes has poorer FER
performance than the degree-6 variable nodes. And the degree-4 variable nodes
contribute almost all the frame errors especially in the low crossover
probability region, which is consistent with the simulation result of the DE
algorithm. Based on the FER performance, we should put information bits into
different layers according to their importance and QoS requirements to reduce
the probability of re-transmission, making good use of the error correction
capability offered by polarization. The coefficients of degree polynomials
still have great influence on the performance, especially on the degree-4
variable nodes. Although the degree-4 variable nodes have worse FER, code D has more degree-6 variable nodes. So the tradeoff between the
performance of lower layer and the code length of higher layer should be taken
into consideration.

\begin{figure}[!t]
	\centering
	\includegraphics[width=0.5\textwidth]{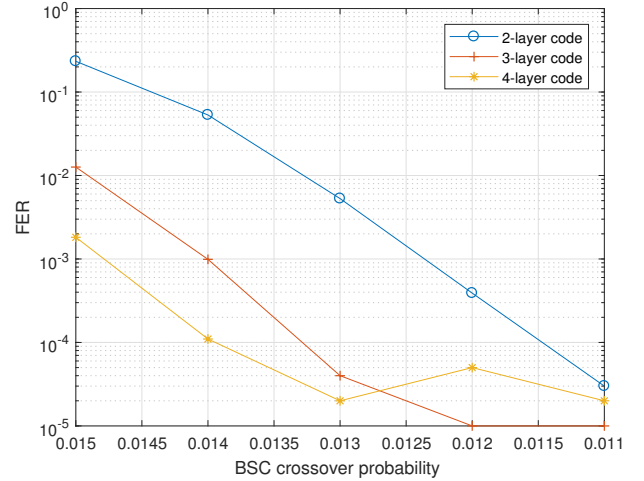}
	\caption{The FER of different Layers of polarized LDPC codes}
	\label{step_FER}
\end{figure}

All 4 codes we considered thus far have two layers (not including degree-2
variable nodes). A natural question is whether the number of layers in the LDPC
codes has influence on the performance. Fig.~\ref{step_FER} shows the FER curves
of LDPC codes having different number of layers. The 2-layer LDPC code has
degree polynomials
\begin{align}
\lambda(x)&=0.0417x+0.8333x^3+0.1250x^5\\
\rho(x)&=0.0417x^{11}+0.8333x^{23}+0.1250x^{35}.
\end{align}
The 3-layer LDPC code has degree polynomials
\begin{align}
\lambda(x)&=0.051x+0.386x^3+0.302x^5+0.261x^7 \\
\rho(x)&=0.065x^{11}+0.351x^{23}+0.404x^{35}+0.180x^{47}.
\end{align}
The 4-layer LDPC code has degree polynomials
\begin{align}
\lambda(x)&\!=\!0.035x\!+\!0.207x^3\!+\!0.310x^5\!+\!0.276x^7\!+\!0.172x^9 \\
\nonumber
\rho(x)&\!=\!0.035x^{11}\!+\!0.207x^{23}\!+\!0.310x^{35}\!\\
&+\!0.276x^{47}\!+\!0.172x^{59}.
\end{align}
In general, the LDPC codes having more layers seems to offer better
performance, especially in the high crossover probability regime. But the
polarized LDPC codes with more layers may have the error propagation problem.
If the higher layer of a codeword has an erroneous bit, the error would
interfere with the decoding of the lower layer bits.

\section{Conclusion}

We proposed a polarized LDPC code design that introduces polarization in the
reliability of the variable and check nodes through judicious connectivity
between the bipartite nodes. The polarized LDPC codes offer a great advantage
in FER on the binary symmetric channel with a slightly increased cost of
decoding complexity per iteration. The polarized LDPC codes could reach a much
lower error-floor which is useful in scenarios where re-transmission of
erroneous frames is costly or impossible, such as in satellite communications.
And the different error correction capability offered by polarization gives
more flexibility to satisfying the different QoS requirements.

\section*{Acknowledgment}
This work was supported by NSF of China Grant No.~61571412 and 
NSF of USA Grant No.~1711922.

\bibliographystyle{IEEEtran}
\bibliography{refs}

\end{document}